\documentclass[fleqn,usenatbib]{mnras}

\usepackage{newtxtext,newtxmath}

\usepackage[T1]{fontenc}

\usepackage{graphicx}	
\usepackage{amsmath}	
\usepackage{color,appendix,soul}

\title[Ultra-massive carbon oxygen white dwarfs]{The evolution of ultra-massive carbon oxygen white dwarfs \thanks{The cooling sequences are publicly available for download at \href{ http://evolgroup.fcaglp.unlp.edu.ar/TRACKS/tracks.html}{http://evolgroup.fcaglp.unlp.edu.ar/TRACKS/tracks.html}}}

\author[Camisassa et al.]{
Mar\'ia E. Camisassa$^{1}$\thanks{E-mail: maria.camisassa@colorado.edu},
Leandro G. Althaus$^{2,3}$,
Detlev Koester$^{4}$,
Santiago Torres$^{5,6}$,
\newauthor 
Pilar Gil Pons$^{5,6}$ and 
Alejandro H. C\'orsico$^{2,3}$
\\
$^{1}$ Department of Applied Mathematics, University of Colorado, Boulder, CO 80309-0526, USA\\
$^{2}$Facultad de Ciencias Astron\'omicas y Geof\'{\i}sicas, 
           Universidad Nacional de La Plata, 
           Paseo del Bosque s/n, 1900 
           La Plata, 
           Argentina\\
$^{3}$ Instituto de Astrof\'isica de La Plata, UNLP-CONICET,
           Paseo del Bosque s/n, 
           1900 La Plata, 
           Argentina\\
$^{4}$ Institut für Theoretische Physik und Astrophysik, Christian-Albrechts-Universität, Kiel 24118, Germany\\           
$^{5}$ Departament de F\'\i sica, 
           Universitat Polit\`ecnica de Catalunya, 
           c/Esteve Terrades 5, 
           08860 Castelldefels, 
           Spain\\
$^{6}$ Institute for Space Studies of Catalonia, 
           c/Gran Capita 2--4, 
           Edif. Nexus 104, 
           08034 Barcelona, 
           Spain\\
}

\date{Accepted XXX. Received YYY; in original form ZZZ}

\pubyear{2015}

\begin{document}
\label{firstpage}
\pagerange{\pageref{firstpage}--\pageref{lastpage}}
\maketitle

\begin{abstract}
Ultra-massive white dwarfs ($\rm M_{WD} \gtrsim 1.05\,  M_{\odot}$) are considered powerful tools to study type Ia supernovae explosions, merger events, the occurrence of physical processes
in the Super Asymptotic Giant Branch (SAGB) phase, and the existence of
high magnetic fields. 
Traditionally, ultra-massive white dwarfs are expected to harbour
oxygen-neon (ONe) cores. However, 
 new observations and recent theoretical studies suggest 
that the progenitors of some ultra-massive white dwarfs can avoid carbon burning, leading to the formation of ultra-massive white dwarfs harbouring carbon-oxygen (CO) cores.
Here we present a set of ultra-massive white dwarf evolutionary sequences with CO cores for a wide range of metallicity and masses. We take into account the energy released by latent heat and phase separation during the crystallization process and by $^{22}$Ne sedimentation. Realistic chemical profiles resulting from the full computation of progenitor evolution are considered. 
We compare our CO ultra-massive white dwarf models with ONe models. We conclude that CO ultra-massive white dwarfs evolve significantly slower than their ONe counterparts mainly for three reasons: their larger thermal content, the effect of crystallization, and the effect of $^{22}$Ne sedimentation. 
We also provide colors in several photometric bands on the basis of new model atmospheres. These CO ultra-massive white dwarf models, together with the ONe ultra-massive white dwarf models, provide an appropriate theoretical framework to study the ultra-massive white dwarf population in our Galaxy. 

\end{abstract}



\section{Introduction} 

White dwarf stars are the most numerous members of the stellar graveyard. It is widely accepted that more
than 95\% of all stars of the Universe, will evolve into white dwarfs, Earth-sized, electron-degenerate objects. Therefore, these numerous objects are considered a powerful tool to 
understand the formation and evolution of stars, the
history of our Galaxy and stellar populations, and the evolution
of planetary systems \citep[see][for reviews]{2008PASP..120.1043F, 2008ARA&A..46..157W,2010A&ARv..18..471A,2016NewAR..72....1G,
2019A&ARv..27....7C}. Among all the white dwarf stars, the ultra-massive white dwarfs, defined as the white dwarfs with masses larger than 
$1.05\, \rm M_{\odot}$, are of special interest. These stars play a key role in our understanding of type Ia Supernova explosions, the occurrence of physical processes in the asymptotic giant-branch (AGB)
phase, the existence of high-field magnetic white dwarfs, and the occurrence of double white dwarf mergers \citep{2015ASPC..493..547D,2020A&A...638A..93R}. The observation of ultra-massive white dwarfs has been reported in several studies in the literature \citep{2004ApJ...607..982M,2016IAUFM..29B.493N,2011ApJ...743..138G,2013ApJS..204....5K,2015MNRAS.450.3966B,2016MNRAS.455.3413K,2017MNRAS.468..239C,2021MNRAS.503.5397K,Hollands2020,2021Natur.595...39C,2021arXiv211009668M}.

The formation of ultra-massive white dwarfs
has been theoretically predicted 
as a result of the isolated evolution of 
massive intermediate-mass stars with a masses larger than $6-9 \, \rm M_\odot$,
depending on the metallicity and the treatment of convective boundaries 
\citep[e.g.][and references therein]{doherty2017}. 
Unlike the progenitors of the less massive white dwarfs, 
the progenitors of ultra-massive white dwarfs
are expected to reach the SAGB with a partially degenerate CO core that reaches temperatures high enough to develop off-center C ignition. 
This violent ignition occurring through flashes is followed by a steady flame propagation towards the centre of the star, and results in the formation of a core
composed mainly by $^{16}$O and $^{20}$Ne, with traces of $^{23}$Na and $^{24}$Mg \citep{1994ApJ...434..306G,2007A&A...476..893S,2006A&A...448..717S,2010A&A...512A..10S}. These stars will eventually lose their outermost layers by the action of stellar winds,
evolving into an ONe-core ultra-massive white dwarf, more massive than
$1.05\, \rm M_{\odot}$. 

During the last years, another scenario for the formation of ultra-massive white dwarfs has gained relevance. Recent studies
have evinced that a relevant fraction of the single white dwarfs
in our Galaxy  have been formed as a result of binary mergers \citep{2017A&A...602A..16T, 2018MNRAS.476.2584M, 2020A&A...636A..31T,2020ApJ...891..160C}. Among these studies, \cite{2020A&A...636A..31T}
estimated, based on binary population synthesis studies, that 
from 30 to 45\% of all the single white dwarfs  with masses larger than 0.9M$_\odot$, in a volume-limited sample, are likely formed through binary mergers, and most of them via double white dwarf mergers. Moreover, \cite{2020ApJ...891..160C} concluded, on the basis of their velocity distribution, that about 20\% of the single white dwarfs with masses larger than 0.8M$_\odot$ are formed through the merger of two white dwarfs. The formation of ultra-massive white dwarfs as a result of double white dwarf mergers has been calculated through hydrodynamical codes in
\cite{2007MNRAS.380..933Y,2009A&A...500.1193L,2014MNRAS.438...14D}. More recently, \cite{2021ApJ...906...53S} studied the merger of two CO core white dwarfs, finding that the initial CO core composition of the merger remnant is converted to ONe, if the remnant has a mass larger than $\sim 1.05$ M$_\odot$. According to this study, the core-chemical composition of an ultra-massive white dwarf formed via mergers should also be ONe.

Nevertheless, recent theoretical models of isolated progenitors of ultra-massive white dwarf suggest that these progenitors can avoid carbon burning on the AGB, supporting the existence of CO-cores in white dwarfs more massive than $1.05\, \rm M_{\odot}$. \cite{ALTUMCO2021}
proposed two possible scenarios for the formation of ultra-massive CO-core white dwarfs in isolated evolution.
One scenario involved a reduction in wind 
rates with respect to the standard prescriptions, in stars with degenerate 
cores of masses lower than $1.05\, \rm M_{\odot}$ at the onset of the 
thermally-pulsing AGB phase. This allowed a slow but significant 
growth of the CO core to values characteristic of CO-core
ultra-massive white dwarfs.
This scenario exploited the lack of understanding of wind rates on first principles and the particularly poor knowledge of
mass-loss in very massive AGB (and SAGB) stars.
The alternative scenario  involved 
 the effects of rotation on the structure of CO cores \citep[see also][]{dominguez1996}, which lowered their internal pressure, 
hampered C-ignition, and ultimately narrowed the initial mass range 
for the formation of ONe WDs. According to \cite{ALTUMCO2021},
 the higher the rotation rates of the progenitor model stars, the larger the
probability of the formation of CO-core ultra-massive white dwarfs. These authors not only 
suggested a feasible scenario for the formation of CO-core ultra-massive white dwarfs,
but also provided accurate core-chemical profiles for these stars.


The existence of CO core ultra-massive white dwarfs is suggested by the analysis of the observations provided by {\it Gaia} space mission of the white dwarf population in the Galaxy \citep{2018A&A...616A..10G}.  
Indeed, the study of \cite{2019ApJ...886..100C} suggests
that 6\% of the ultra-massive white dwarf population should have an extra-delay of nearly 8 Gyrs on their cooling times. Several authors have proposed solutions to this extra-cooling anomaly on the ultra-massive white dwarfs \citep{2020ApJ...902...93B,2020PhRvD.102h3031H,2021ApJ...919L..12C,2021ApJ...911L...5B, 2021A&A...649L...7C} but none of them has lead to conclusive results. Moreover, the most  promising solutions to the extra-cooling anomaly invoke that the ultra-massive white dwarfs that experience the extra-cooling delays harbour CO cores.

Although the evolution of ultra-massive ONe-core white dwarfs has been analyzed in detail in \cite{camisassa2019}, 
 there are few works in the literature devoted to the study of the evolution ultra-massive CO-core white dwarfs. \cite{2021ApJ...916..119S} calculated ultra-massive CO white dwarfs, but focused on the most massive white dwarfs, with masses larger than $1.3\,\rm M_\odot $. On the other hand, \cite{Bedard2020} provided ultra-massive CO white dwarf models, but these authors considered artificial initial chemical profiles composed by
a constant profile of 50\% O and 50\% C. The accurate core-chemical composition is a key ingredient in obtaining reliable cooling times.
Furthermore, these calculations did not include the energy released due to the phase separation process induced by crystallization, nor the energy 
released 
by $^{22}$Ne sedimentation. The recent studies of \cite{2021A&A...649L...7C} have shown that, although the effect of $^{22}$Ne sedimentation is negligible in the evolution of ONe ultra-massive white dwarfs, it is crucial to determine accurate cooling times for CO ultra-massive white dwarfs. This paper is precisely aimed at filling this theoretical gap, providing a set of ultra-massive CO-core white dwarf evolutionary models that include all the relevant energy sources and an accurate chemical profile that results of the full progenitor evolution. Indeed, we consider the initial chemical profiles of \cite{ALTUMCO2021}.
We also include the energy released 
by $^{22}\rm Ne$ sedimentation, by latent heat and due to the core chemical redistribution during the crystallization process, 
following the phase diagram of \cite{2010PhRvL.104w1101H}, suitable for C and O composition. 
We also provide accurate magnitudes and colors for
our models in the filters used by the spacial mission {\it Gaia}: G, $\rm G_{BP}$, and $\rm G_{RP}$ and by the Sloan Digital Sky Survey (SDSS): $u$, $g$, $r$, $i$ and $z$, using model atmospheres of \cite{2010MmSAI..81..921K,2019A&A...628A.102K}.
To the best of our knowledge, this is the first set of fully
evolutionary calculations of ultra-massive CO-core white dwarfs including a realistic initial chemical profile, an updated microphysics and the effects of the crystallization process and $^{22}$Ne sedimentation.
This set of cooling sequences, together with the ONe ultra-massive models of \cite{camisassa2019}, provide an appropriate theoretical framework to study the ultra-massive white dwarf population in our Galaxy. 

The paper is organized as follows. In Section \ref{LPCODE}  we describe the theoretical modelling of 
our evolutionary calculations. That includes a description of the microphysics considered, the treatment of the energy released by the different energy sources and the chemical profiles considered.
In Section \ref{results} we present the results of our evolutionary calculations and we compare them with other ultra-massive white dwarf cooling models existing in the literature. 
Finally, in Section \ref{conclusions} we analyze the results achieved and summarize them in the concluding remarks.

\section{Theoretical white dwarf modelling}
\label{LPCODE} 

The white dwarf evolutionary models presented in this work
were calculated using the {\tt LPCODE} stellar evolutionary code \citep[see][for details]{2005A&A...435..631A,2015A&A...576A...9A}. This evolutionary code
has been amply used in the study of different aspects of low-mass star evolution \citep[see][and references therein]{2010ApJ...717..897A,2010ApJ...717..183R}. Moreover, the code has been recently used to generate a new
grid of models for post-AGB stars \citep{2016A&A...588A..25M} and
new white dwarf evolutionary sequences \citep{2016ApJ...823..158C,2017ApJ...839...11C,camisassa2019,2021A&A...649L...7C}. This code 
has been tested and calibrated against other codes
in different evolutionary phases, such
as the red giant phase \citep{2020A&A...635A.164S,2020A&A...635A.165C}
 and the white dwarf phase \citep{2013A&A...555A..96S}, 
We briefly summarize the main input physics of
{\tt LPCODE}
 for the white dwarf regime.
The radiative and conductive opacities are taken from OPAL \citep{1996ApJ...464..943I} and from \cite{2007ApJ...661.1094C}, respectively. For the low-temperature
regime, the code considers molecular radiative opacities with varying carbon-to-oxygen ratios, computed by \cite{2005ApJ...623..585F}. 
The code treats convection
within the standard mixing length formulation, as given by the
ML2 parameterization \citep{1990ApJS...72..335T}. 
We considered the
neutrino emission rates of \cite{1996ApJS..102..411I} for pair, photo, and bremsstrahlung processes, whereas for
plasma processes
we follow the prescription from \cite{1994ApJ...425..222H}. The outer
boundary conditions of our evolving models are provided by non-gray model atmospheres both for H-rich and H-deficient composition \citep[see][for references]{2012A&A...546A.119R, 2017ApJ...839...11C,2018MNRAS.473..457R}. Finally, {\tt LPCODE} takes into account the
changes in the chemical abundances due to 
element diffusion, including gravitational settling, and chemical and thermal diffusion.

The initial white dwarf models are result of the full progenitor evolution of the model stars computed in 
\cite{ALTUMCO2021}, with the Monash-Mount Stromlo code
 {\tt MONSTAR} \citep{wood1987, frost1996, campbell2008,gilpons2013, gilpons2018}. 
In this work, we consider the chemical profile that results from the scenario in which the authors computed the evolution 
of a 7.8 M$_\odot$ model, from the Zero Age Main Sequence,
until advanced stages of the thermally-pulsing AGB, when the CO 
core-mass grew above 1.05 M$_\odot$. Note that this evolutionary sequence was computed scaling-down the mass-loss rates of
\cite{vassiliadis1993} along the early AGB and the 
thermally-pulsing AGB. 
Therefore, in our white dwarf models, the C/O proportion in the core and the total He mass is the result of the full progenitor evolution. The H mass is set to $\sim 10^{-6}\, \rm M_\odot$. The core-chemical profile obtained by these authors in this scenario does not present large discrepancies with the one obtained in their so-called stellar rotation scenario. The main differences in the final chemical profiles relies in the mass of the He mantle. That is, the
rotation scenario predicts larger He masses that will ultimately be burned at the beginning of the cooling sequence. Therefore, we chose to
employ the chemical profile that results of the reduced mass loss scenario, relying on its most accurate determination of the total He mass \citep[see][for details]{ALTUMCO2021}.

 Although \cite{ALTUMCO2021} explored the evolutionary properties of carbon oxygen 
ultra-massive
white dwarfs, these authors only performed evolutionary calculations for $1.156\, \rm M_\odot$ models, and did not consider the energy released by 
$^{22}\rm Ne$ sedimentation, since that paper was aimed at providing a plausible
scenario for the formation of ultra-massive white dwarfs with carbon oxygen cores. In our paper we consider four white dwarf masses, $1.10\, \rm M_\odot$, 
 $1.16\, \rm M_\odot$, $1.23\, \rm M_\odot$ and $1.29\, \rm M_\odot$, to be consistent with the ONe cooling sequences of \cite{camisassa2019}. 
For each mass, we calculate 8 cooling sequences, one H-rich and one H-deficient for each of the following initial metallicities $Z=0.001$, $Z=0.02$, $Z=0.04$ and $Z=0.06$. In each cooling sequence, we artificially set the initial $^{22}$Ne abundances to be constant through the core and equal to the inital metallicity.

 At the beginning of the cooling track we perform a mixing of all core chemical components
 in some unstable layers of the core. This mixing is expected to occur as a result of the inversion of the mean molecular weight
in the CO core. We simulate this process by assuming the mixing to be instantaneous. The chemical abundances will change during the white dwarf evolution due to the action of element diffusion. Gravitational settling acts in purifying the envelopes of our models, leaving  pure-H envelopes for our H-rich models, and pure He envelopes for our H-deficient models.
 Chemical and thermal diffusion act as smoothing across chemical interfaces.

\subsection{Crystallization}
\label{ss:crys} 

The evolutionary models presented in this work  consider both the energy released as latent heat and by the phase separation of carbon and oxygen during the crystallization process.
We employed the phase diagram of \cite{2010PhRvL.104w1101H} to obtain the
crystallization temperatures and the changes in the chemical abundances due to crystallization. The new CO phase diagram of \cite{2020A&A...640L..11B} yields similar results, and the azeotropic composition is not reached in our models. 
We have not included  $^{22}$Ne distillation process \citep{2021ApJ...911L...5B}, a process whose occurrence is under debate,
because, to date, there is not a complete phase diagram published in the literature that covers the entire range of C-O-Ne abundances.
The energy released by the crystallization process is included self consistently and locally coupled to the full set of
equations of stellar evolution. We considered a two-component crystallization process, where the white dwarf interior is made only of carbon and oxygen with abundance by mass $X_{\rm C}$ and $X_{\rm O}$ respectively
($X_{\rm C} + X_{\rm O}=1$). Then, the luminosity equation considered is \citep[see][]{2008ApJ...677..473G,2010ApJ...719..612A}):

\begin{equation}
\label{eq:lum}
\frac{\partial L_r}{\partial M_r} = \epsilon_{\rm nuc}-\epsilon_{\nu}-C_P \frac{dT}{dt}+\frac{\delta}{\rho} \frac{dP}{dt}+ l_s \frac{dM_s}{dt} \delta(m-M_s)-A\frac{dX_{\rm O}}{dt}
\end{equation}
where $A$ is given by:
\begin{equation}
A= \left(\frac{\partial u}{\partial X_{\rm O}}\right)_{\rho,T}+ \frac{\delta}{\rho} \left( \frac{\partial P}{\partial X_{\rm O}} \right)_{\rho,T} \approx \left(\frac{\partial u}{\partial X_{\rm O}}\right)_{\rho,T}
\end{equation}
and $u$ is the internal energy per gram.

In equation \ref{eq:lum}$, \epsilon_{\rm nuc}$ is the energy released by nuclear reactions, which is disregarded in our white dwarf calculations, and $\epsilon_\nu$ is the
energy released by neutrino emission. The third and fourth term in equation \ref{eq:lum} correspond to the gravothermal energy release.
The fifth term denotes the energy released
as latent heat, where $l_s$ is the latent heat of crystallization and
$\frac{dM_s}{dt}$ is the rate at which the solid core grows. The
delta function indicates that the latent heat is released
at the solidification front. 
 We considered this latent heat release to be $0.77\,k_B\, T$ per crystallized ion, where $T$ is the temperature at the crystallization front.
Finally, the last term corresponds to the energy released by 
changes in the chemical abundances. This term is usually neglected in normal stars, since it is much smaller than the energy released
by nuclear reactions. However, it plays a major role in crystallizing white dwarfs, due to the phase separation process. In the case of
CO white dwarfs, $\left(\frac{\partial u}{\partial X_{\rm O}}\right)_{\rho,T}$ is dominated by
the ionic contributions, and is negative. Therefore, the last
term in Equation \ref{eq:lum} will be a source of energy in
those regions where the oxygen abundance increases and a sink of energy in those regions where the carbon abundance increases. Following the
phase diagram of \cite{2010PhRvL.104w1101H} for a CO
white dwarf, the oxygen abundance in the crystallizing
region increases, and the surrounding liquid region is depleted in oxygen.
This induces a mixing process at the region above the
crystallized core due to a Rayleigh-Taylor instability. 
Therefore, the crystallization process will lead to a source of energy in
the crystallizing regions, and a sink of energy in the layers above it, which are depleted on oxygen.

  When crystallization sets in, a Rayleigh-Taylor instability rapidly induces a convective region that will characterize nearly 70\% of the white dwarf mass, depending on the initial white dwarf chemical profile. We compute the resulting chemical rehomogenization assuming this mixing to be instantaneous \citep{1997ApJ...485..308I}.
 As the evolution proceeds, the crystallization front moves outward and so does the inner boundary of the mixing region, while its outer boundary barely moves outward. \cite{2017ApJ...836L..28I} suggested that this convective mixing can produce
a dynamo able to yield magnetic fields as they occur in Solar System planets.

To avoid numerical difficulties when integrating the full set of equations, we computed the crystallization energy released integrating the last term in equation \ref{eq:lum} over the whole star during the evolutionary time-step \citep{1997ApJ...485..308I}. Therefore, the energy released by the phase separation process
during the evolutionary time-step is:

\begin{equation}
\label{eq:dudo}
\int_0^M \left(\frac{\partial u}{\partial X_{\rm O}}\right) \frac{dX_{\rm O}}{dt} dM_r= (X_{\rm O}^{\rm sol}-X_{\rm O}^{\rm liq}) \times
\left[ \left(\frac{\partial u}{\partial X_{\rm O}}\right)_{M_s}-\left<\frac{\partial u}{\partial X_{\rm O}}\right>
\right]\frac{dM_s}{dt}
\end{equation}
 where $\left(\frac{\partial u}{\partial X_{\rm O}}\right)_{M_s}$ is evaluated at the crystallization boundary and
\begin{equation}
\left<\frac{\partial u}{\partial X_{\rm O}}\right> =
\frac{1}{\Delta M} \int_{\Delta M} \left(\frac{\partial u}{\partial X_{\rm O}}\right) dM_r
\end{equation}

$X_{\rm O}^{\rm sol}$ and $X_{\rm O}^{\rm liq}$ are the solid and liquid abundances of oxygen at the crystallization front. The first term in the square bracket in Equation \ref{eq:dudo} corresponds to
the energy released at the crystallization front, and
the second term, is the energy absorbed, on average, on the convective region ($\Delta M$) driven
by the Rayleigh-Taylor instability. Note that the 
exact initial chemical profile is crucial to asses the mixing in the liquid 
layers above the crystallization front, and therefore, the energy absorbed there. Thus, the exact shape of the chemical profile is crucial in determining  the net energy released by the crystallization process.

\subsection{$^{22}\rm Ne$ sedimentation}
\label{ss:ne22}

$^{22}$Ne is the
most abundant trace present in the CO core white dwarfs.
The $^{22}$Ne abundance in the core is roughly equal to the initial metallicity of the star. This abundance is
built up during the helium burning on $^{14}$N via the reactions
$\rm ^{14}N(\alpha,\gamma)^{18}F(\beta+)^{18}O(\alpha,\gamma)^{22}Ne$.
The neutron excess in $^{22}$Ne with respect to the other components of the white dwarf interior, which have equal number of protons and neutrons, results in a net downward gravitational force, inducing
the slow settling of $^{22}$Ne in the liquid regions toward the center of
the white dwarf \citep{2002ApJ...580.1077D,2010ApJ...719..612A,2016ApJ...823..158C}.
 To compute the energy released due to $^{22}$Ne sedimentation, we add a last term to Equation \ref{eq:lum},  that corresponds to the energy released by  the changes in the abundance of $^{22}$Ne. This extra term is \citep[see][for details]{2008ApJ...677..473G}:
\begin{equation}
-\left[ \left( \frac{\partial u}{\partial X_{\rm Ne22}} \right)_{\rho,T} - \frac{\delta}{\rho} \left( \frac{\partial P}{\partial X_{\rm Ne22}}
\right) \right ] \frac{dX_{\rm Ne22}}{dt}
\end{equation}
Since $^{22}$Ne is a neutron-rich isotope, the derivative $\left( \frac{\partial u}{\partial X_{\rm Ne22}} \right)_{\rho,T}$ is dominated by the
electronic contributions. This is, an increase (decrease) in the local
$^{22}$Ne abundance implies an increase (decrease) in the
molecular weight per electron $\mu_e$, forcing the degenerate electron gas to release (absorb) energy.
Therefore, $^{22}$Ne sedimentation will be a source (sink) of energy
in the regions where the local $^{22}$Ne abundance increases (decreases). 

To compute this process, we employed the diffusion coefficients of \cite{2010PhRvE..82f6401H}. We assumed that $^{22}$Ne diffuses in a one-component plasma background consisting of a fictitious element with atomic weight ($A$) and atomic number ($Z$), defined by the average $A$ and $Z$ in each
layer and not assumed constant through different layers \citep[see][for details]{2016ApJ...823..158C}. During the crystallization process, the liquid regions are carbon enriched, and this carbon enrichment is taken into account when calculating the $^{22}$Ne sedimentation. The diffusion of $^{22}$Ne is more efficient for a lower $Z$ of the background element and, therefore,  $^{22}$Ne sedimentation in the liquid regions will be encouraged by the carbon enrichment in these regions.

\section{Results}
\label{results} 

\begin{figure}
      {\includegraphics[width=1.0\columnwidth, clip=true,trim=0 0 0 0]{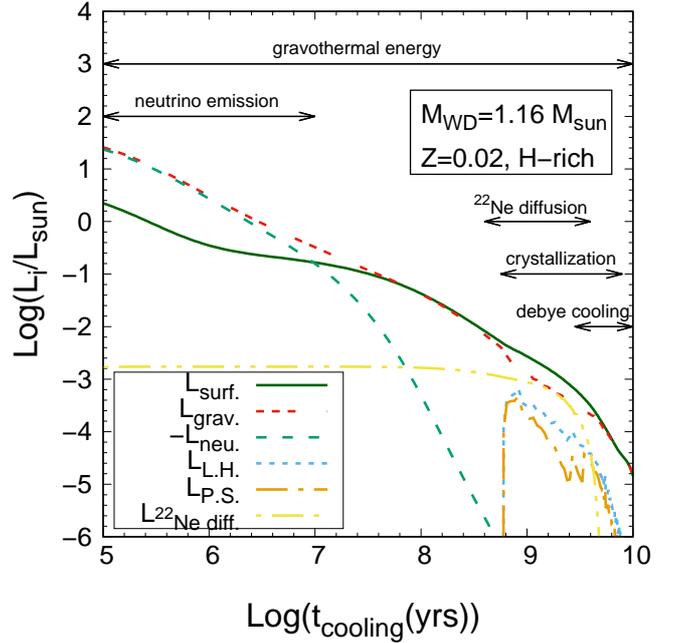}}
  \caption{Main contributions to the white dwarf luminosity (solid dark-green line) in terms of
  the cooling times for our 1.16M$_\odot$ H-rich model with $Z=0.02$ . We plot the gravothermal luminosity (triple-dashed red line), the energy lost by
  neutrino emission (dashed green line), the energy released as latent heat (dotted cyan line), the energy released due to phase separation upon crystallization (dot-dashed orange line), and the energy released by 
  $^{22}$Ne sedimentation (dot-dot-dashed yellow line).
The arrows indicate the
main physical processes responsible for the evolution at the different stages of the white dwarf cooling.} 
  \label{f:lumi}
\end{figure}

\begin{figure}
      {\includegraphics[width=1.0\columnwidth, clip=true,trim=0 0 0 0]{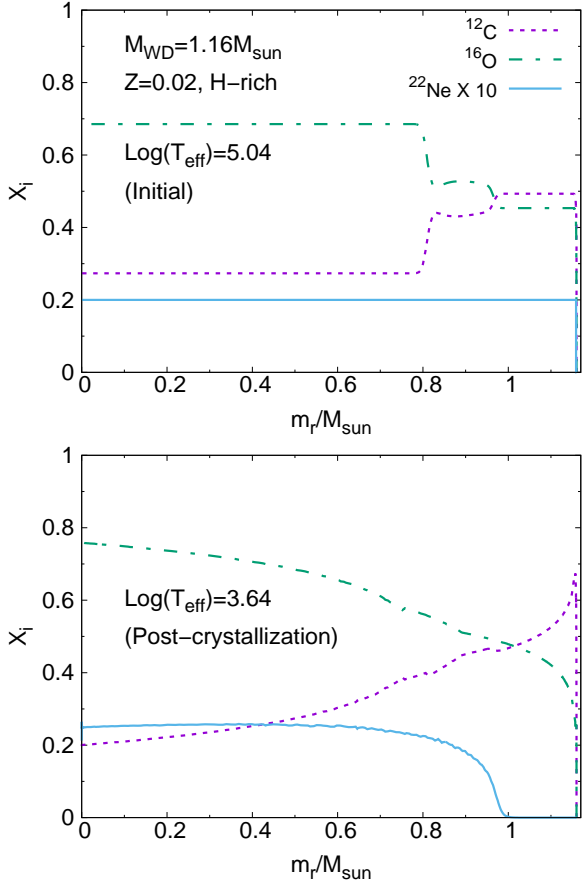}}
  \caption{Initial (top panel) and final (bottom panel)
  chemical profile of our 1.16M$_\odot$ H-rich ultra-massive white dwarf model with $Z=0.02$. The abundances of $^{12}$C, $^{16}$O and $^{22}$Ne are shown as dotted,  dot-dashed and solid lines, respectively. Note that the $^{22}$Ne abundance is multiplied by a factor 10, to make it more visible in the plot.  } 
  \label{f:profile}
\end{figure}

In Fig. \ref{f:lumi} we show  a global view of the main energy contributions to the evolution of a typical ultra-massive 
CO white dwarf model. This Figure shows the different luminosity contributions in our 1.16M$_\odot$
H-rich white dwarf model with solar $^{22}$Ne abundance ($Z=0.02$ model)
in terms of the logarithm of the white dwarf cooling time. We define the white dwarf cooling time as the time since the white dwarf reached the maximum effective temperature. The contribution of all the different sources and sinks of energy is equal to the total luminosity. During the entire white dwarf cooling phase,
the gravothermal energy is the main energy contribution to the stellar
luminosity. At early stages of the white dwarf cooling, the energy released by neutrino emission is an important sink of energy, as it is larger than the stellar luminosity. However, as the evolution proceeds, the white dwarf cools down and the neutrino emission decreases until it is overpassed by the white dwarf luminosity when the cooling age is $\sim 10^7$ years. The energy released by the slow settling of $^{22}$Ne is
nearly constant until it rapidly drops due to the white dwarf crystallization. Therefore, the energy released by  $^{22}$Ne sedimentation increases its contribution to the total luminosity as the gravothermal energy decreases. It is important to remark that, as the crystallization sets in in the inner regions of the core, $^{22}$Ne continues its settling in the outer liquid regions, and this settling is encouraged by the C enhancement in the liquid regions.
When the surface luminosity reaches $10^{-2.34}\rm \, L_\odot$, crystallization starts in the center of our $1.16 \rm \, M_\odot$ model, when the cooling age is about $0.612\times 10^9$ years.
This process releases latent heat and gravitational energy due to the
chemical redistribution of C and O. Although this process occurs at lower luminosities than in an ONe ultra-massive white dwarf with the same mass \citep[see Fig 4. of][]{camisassa2019}, it still occurs when the gravitational energy release is large, thus diminishing its effect on the cooling times.
Finally, by the end of the cooling sequence,
the temperature of the core drops below
the Debye temperature, and consequently, the heat capacity is substantially reduced. Therefore, the white dwarf enters a rapid cooling phase, called the Debye cooling phase.

The initial and final chemical profiles of this white dwarf cooling sequence are shown in Figure \ref{f:profile}. The abundances of the most important isotopes in this white dwarf model interior, $^{12}$C, $^{16}$O and $^{22}$Ne, are shown. Note that the $^{22}$Ne abundance is multiplied by a factor 10, for visualization purposes. The initial chemical profile is the result of the progenitor evolution through the AGB calculated in \cite{ALTUMCO2021}. The original chemical profile has been homogenized due to Rayleigh Taylor instabilities in the initial white dwarf composition (see Section \ref{LPCODE}). 
In the bottom panel of Figure \ref{f:profile}, we show the final chemical profile that results of the crystallization process and $^{22}$Ne sedimentation. At this stage, more than 99\% of the white dwarf model has crystallized. Note the $^{16}$O enrichment in the innermost layers and $^{12}$C enrichment in the outermost layers, as a result of phase separation due to crystallization. Since  $^{22}$Ne 
diffusion coefficient is larger for a lower atomic number $Z$ of the background, the $^{12}$C enrichment in the outermost layers encourages  $^{22}$Ne
sedimentation in the liquid outer regions when the inner regions are crystallized. Therefore,  $^{22}$Ne
sedimentation 
has to be calculated consistently with the crystallization process to provide accurate cooling times for white dwarfs, specially when calculating models with high $^{22}$Ne abundances.

\subsection{The ultra-massive white dwarf cooling times}

\begin{figure*}
      {\includegraphics[width=1.0\columnwidth, clip=true,trim=0 0 0 0]{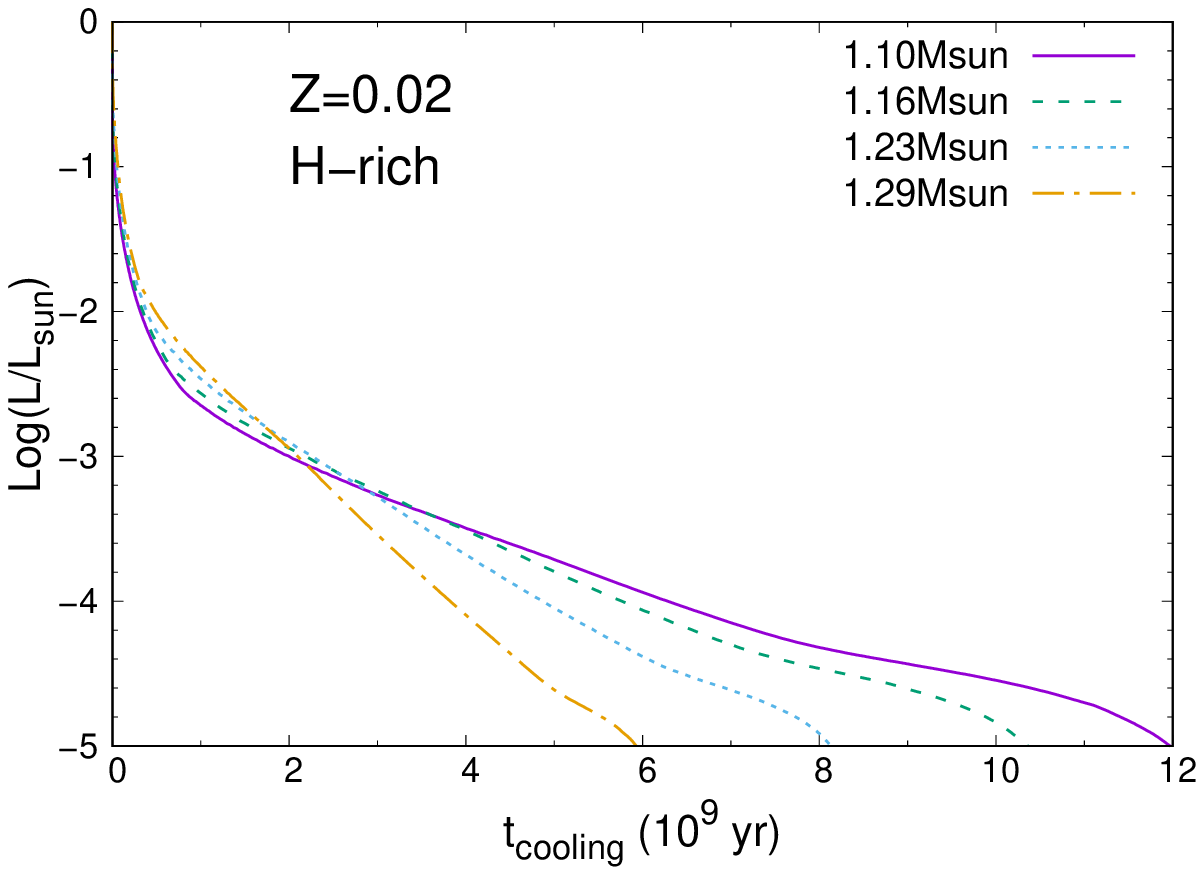}}
        {\includegraphics[width=1.0\columnwidth,clip=true,trim=0 0 0 0]{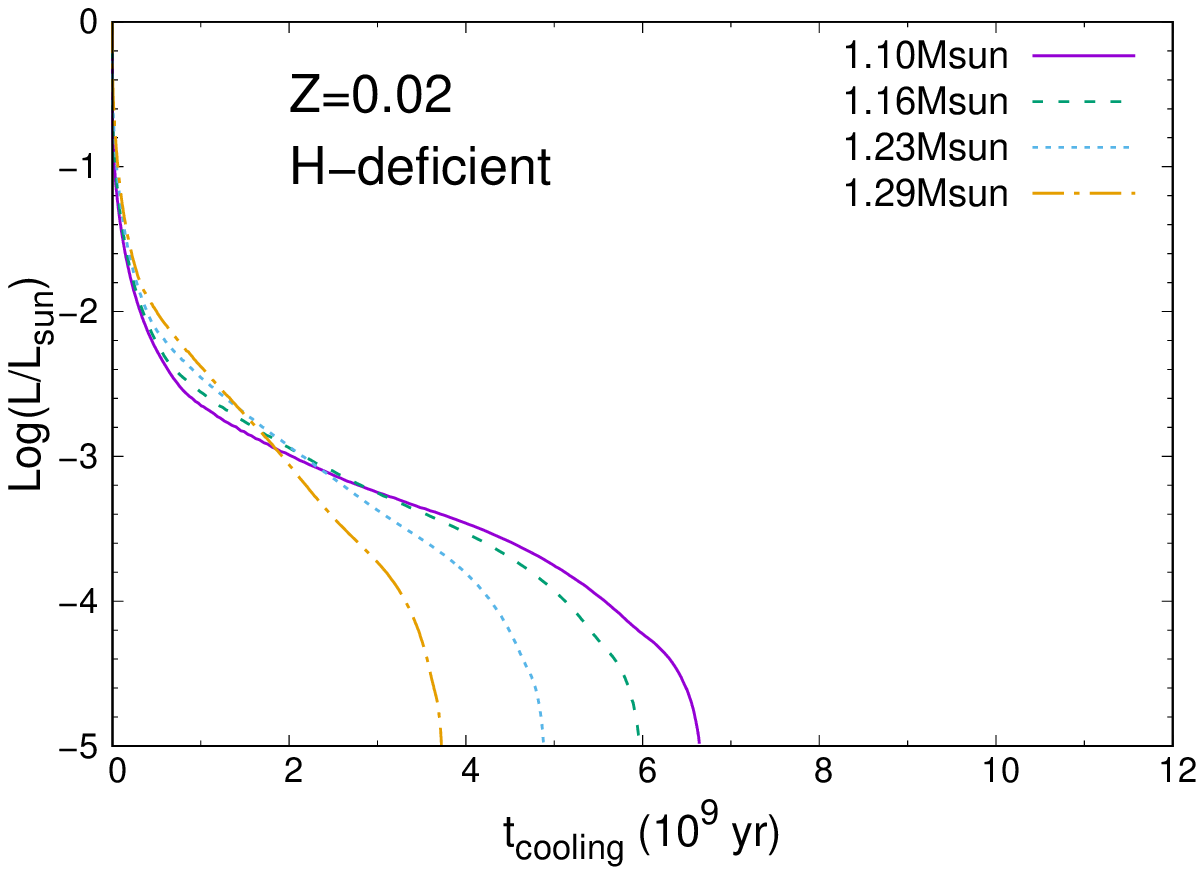}}  
  \caption{Left panel: Cooling times of all our H-rich evolutionary models with metallicity $Z=0.02$. Right panel: Same as left panel, but for our H-deficient evolutionary models with metallicity $Z=0.02$.} 
    \label{f:evol_all}
\end{figure*}

\begin{figure}
      {\includegraphics[width=1.0\columnwidth, clip=true,trim=0 0 0 0]{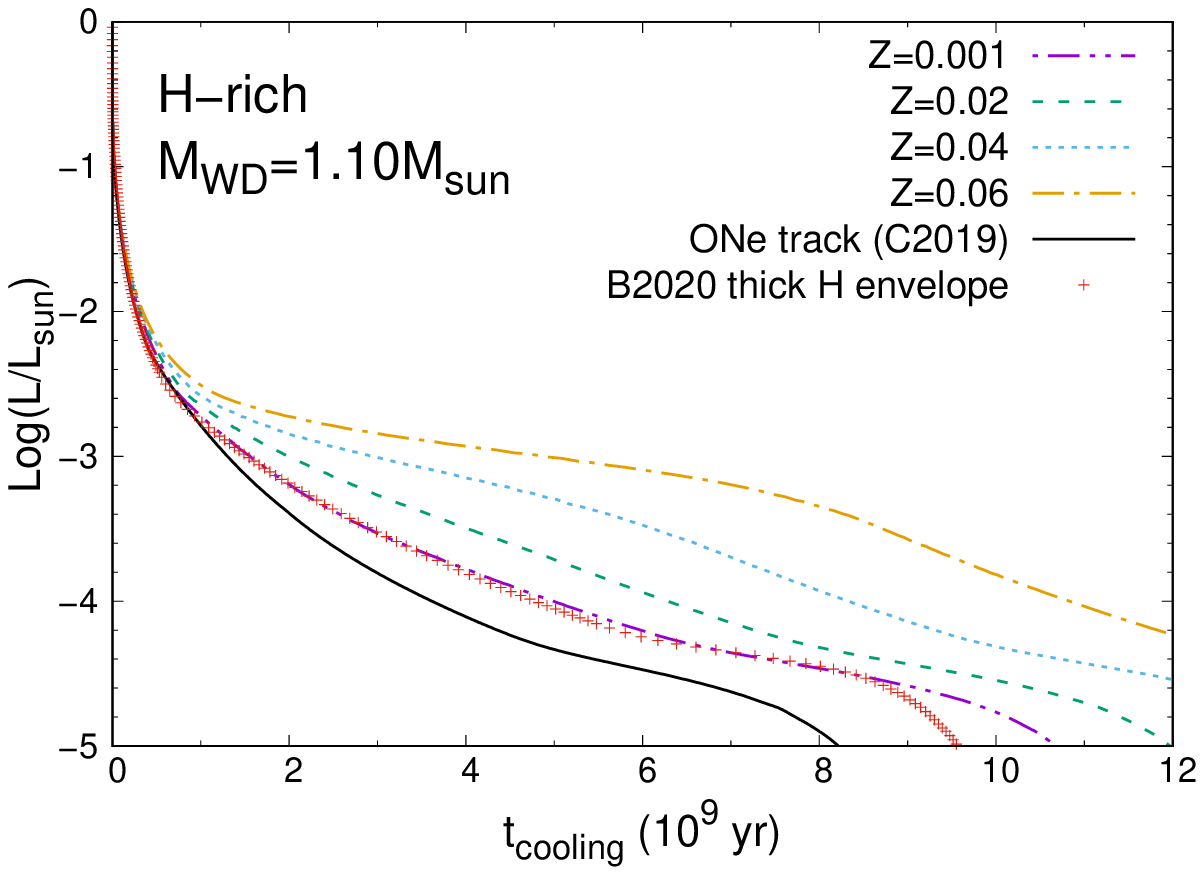}}
            {\includegraphics[width=1.0\columnwidth, clip=true,trim=0 0 0 0]{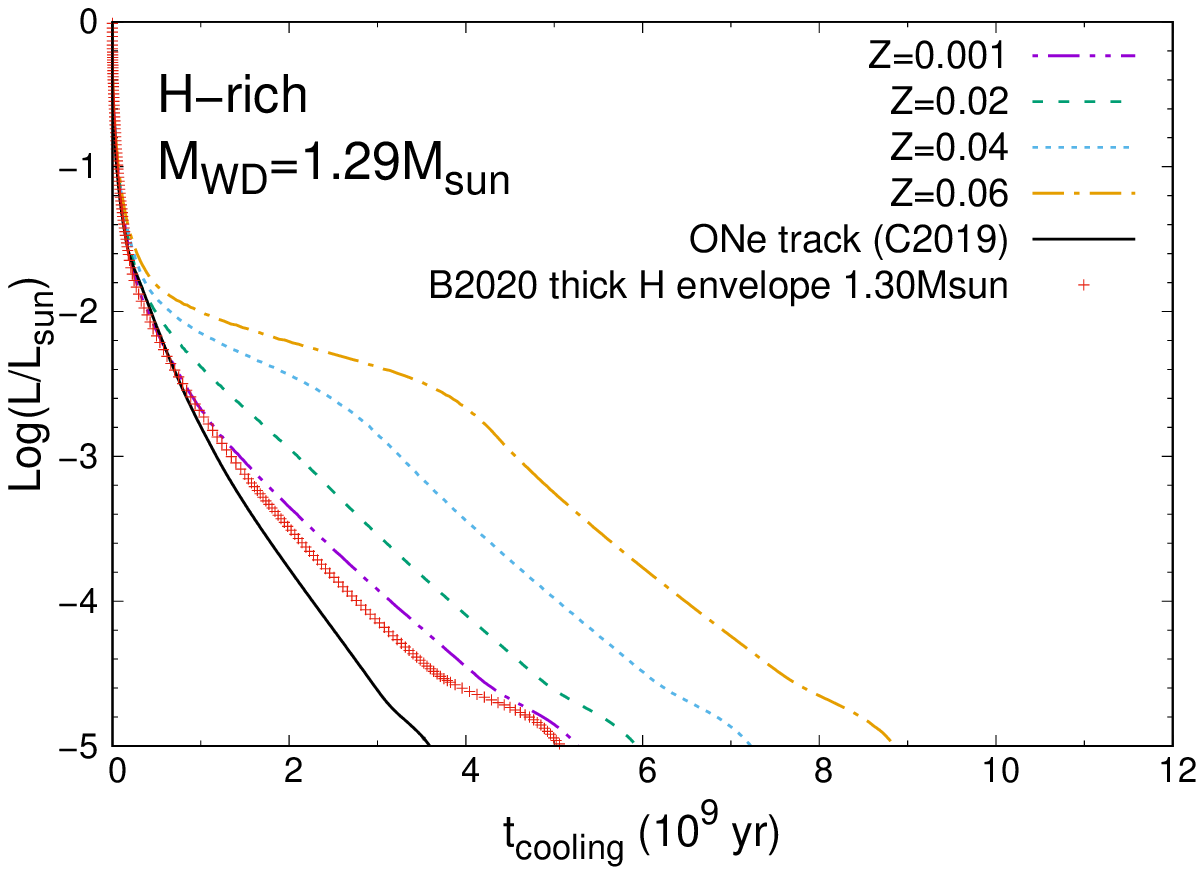}}
  \caption{Top panel: Cooling times of our CO-core 1.10 $\rm M_{\odot}$ ultra-massive white dwarf models with H envelopes for different metallicities (see text for details), together with the ONe-core model of \protect \cite{camisassa2019} with H-envelope and the CO-core model with thick H envelope of \protect \cite{Bedard2020} 
  with the same mass.  Bottom panel: Same as top panel,
  but for our 1.29 $\rm M_{\odot}$ ultra-massive white
  dwarf models with H envelopes. Note that the CO-core
  model with thick H envelope of \protect \cite{Bedard2020} has
  1.30 $\rm M_{\odot}$, being 0.01 $\rm M_{\odot}$ more
  massive than the rest of the models displayed in this
  panel.}
    \label{f:ageZ}
\end{figure}

One of the main goals of this study is to provide accurate cooling times for ultra-massive white dwarf models with CO cores. In Figure \ref{f:evol_all} we show the luminosity as a function of the cooling times for all our solar metallicity (Z=0.02) models. The models with a H (He) envelope are shown in the left (right) panel of this Figure. 
The differences in the cooling
behavior for different white dwarf masses result clear by inspecting this figure. The most massive white dwarfs evolve slower at higher luminosities, but this trend is reversed at lower luminosities, where the more massive the white dwarf is, the faster its cooling is. The short cooling times found for the most
massive sequences relies on the fact that these stars enter the Debye regime at higher luminosities, due to their larger characteristic densities, and the
specific heat of the ions is strongly reduced. In particular, note the fast cooling of the H-rich 1.29M$_\odot$ white dwarf model, that reaches 
$\log(L/\rm L_\odot)\sim -5$ in $5.9\times 10^9$ years.
Moreover, the H-deficient 1.29M$_\odot$ white dwarf 
sequence reaches $\log(L/\rm L_\odot)\sim -5$ in less than $3.72\times 10^9$ years.
In general lines, all our H-deficient CO ultra-massive white dwarf models evolve faster than their H-rich counterparts. This result is
 consistent with previous studies of hydrogen-deficient white dwarfs of \cite{2017ApJ...839...11C,camisassa2019}.
The faster cooling experienced by H-deficient white dwarfs is mainly due to two reasons. First, these stars held outer layers that are more transparent to radiation, and second, the convective coupling occurs earlier in these stars, with the subsequent energy release at higher luminosities. 
 This Figure can be compared with Figure 5 of \cite{camisassa2019}, where the authors find a similar trend for ONe ultra-massive white dwarfs.
Finally, the models with other initial metallicities (Z=0.001, Z=0.04 and Z=0.06) calculated in this work are not shown on this plot, but they present a similar behaviour. 

The effect on the cooling times due to the different initial metallicities (or $^{22}$Ne abundances) is shown in Figure \ref{f:ageZ}, together with the comparison between our cooling times and the cooling times found in \cite{camisassa2019} and \cite{Bedard2020}.
The top panel of this Figure shows the luminosity in terms of the cooling age for our H-rich 1.10M$_\odot$ models with $Z=0.001$, $Z=0.02$, $Z=0.04$ and $Z=0.06$.
The H-rich ONe evolutionary sequence of \cite{camisassa2019} and the thick H-envelope model of \cite{Bedard2020}
with the same mass are shown as a solid black line and as a red line, respectively.
In the bottom panel of this Figure we plot the luminosity in terms of the cooling age for our H-rich 1.29M$_\odot$ models with $Z=0.001$, $Z=0.02$, $Z=0.04$ and $Z=0.06$, together with the ONe evolutionary sequence of \cite{camisassa2019} with the same mass and the thick H-envelope model of \cite{Bedard2020} with 1.30M$_\odot$.
A first glance of this Figure reveals that ultra-massive CO white dwarfs evolve significantly slower than their ONe counterparts. This result was already noticed in \cite{2021A&A...649L...7C}, and still holds for ONe models that include $^{22}$Ne sedimentation, since this process does not impact the evolutionary times of ONe white dwarfs. Note that, even when compared with our $Z=0.001$
models, where the effect of $^{22}$Ne sedimentation is not significant, ONe models evolve markedly faster. The reason for this behaviour relies on the 
lower thermal content that ONe white dwarfs have, since the
specific heat per gram is lower in an ONe plasma than in an CO plasma. Furthermore, ONe white dwarfs crystallize at higher luminosities than CO white dwarfs with the same mass, and therefore, the impact of crystallization on their cooling times is lower.

From both panels of Figure \ref{f:ageZ} it results clear that, the higher the metallicity (or $^{22}$Ne abundance), the longer the cooling times. 
Since a higher $^{22}$Ne abundance results in a larger energy release by $^{22}$Ne sedimentation, longer cooling times are obtained for models with high initial $Z$. This is particularly true for the 1.10M$_\odot$ cooling sequence with a $^{22}$Ne abundance of 0.06, that would reach the faint end of the cooling sequence ($\log(L/\rm L_\odot) \sim -5$) in a time longer than the age of the universe.
The long cooling times obtained for this sequence are in line with the predictions of \cite{2019ApJ...886..100C} for the cooling anomaly of ultra-massive white dwarfs. The ultra-massive CO white dwarfs with initial metallicity higher than $\sim 0.04$
experience a strong delay on their cooling times due to $^{22}$Ne sedimentation. Therefore, these stars are called "forever young white dwarfs" in \cite{2021A&A...649L...7C}.
If we consider a Z distribution of \cite{2011A&A...530A.138C}, the percentage of white dwarfs expected to have $Z>0.04$ is about 4\%. Therefore, 4\% of the ultra-massive CO white dwarfs should experience a strong delay on their cooling times due to $^{22}$Ne sedimentation.

Finally, the models of \cite{Bedard2020}
evolve slightly faster than our models with $^{22}$Ne abundance 0.001 with the same mass. These authors did not consider the energy released by $^{22}$Ne sedimentation, and therefore, their cooling curves resemble our cooling curves in which the energy released by $^{22}$Ne sedimentation is negligible. Further, these authors 
did not consider the energy released by the phase separation induced by crystallization, 
and thus, their models evolve even faster. The differences in the chemical composition between our models and \cite{Bedard2020} will also induce differences in the cooling times, but the quantification of these differences is beyond the scope of this paper.

\begin{figure}
      {\includegraphics[width=1.0\columnwidth, clip=true,trim=0 0 0 0]{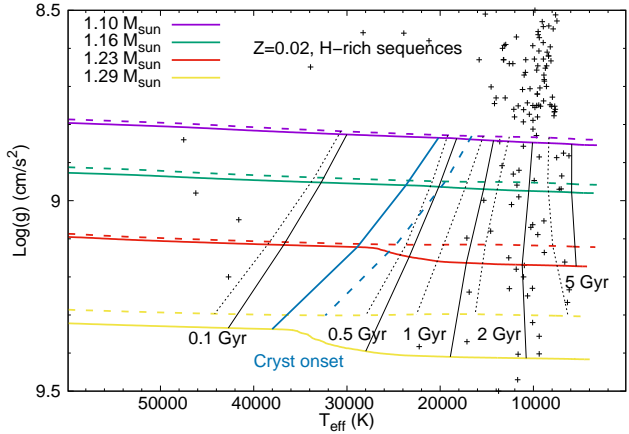}}
  \caption{CO H-rich models (dashed lines) with $Z=0.02$, together with the ONe H-rich models of \protect \cite{camisassa2019} (solid lines) on the $\log \rm g - T_{eff}$ plane. From left to right, the cooling age isochrones of $0.1$, $0.5$, $1$, $2$ and $5$ Gyrs ($10^9$years) in the CO-core (ONe-core) models are shown as black dashed (solid) lines.
   The light-blue dashed (solid) line indicates the crystallization onset for CO (ONe) white dwarf models. Crosses indicate the
location of ultra-massive white dwarfs observed by  \protect \cite{2004ApJ...607..982M,2016IAUFM..29B.493N,2011ApJ...743..138G,2013ApJS..204....5K,2015MNRAS.450.3966B,2016MNRAS.455.3413K,2017MNRAS.468..239C,Hollands2020,2021arXiv211009668M}.} 
  \label{f:loggteff}
\end{figure}

The mass-radius relations of our CO H-rich models and the ONe H-rich models of \cite{camisassa2019} are shown on the $\log \rm g - T_{eff}$ plane in Figure \ref{f:loggteff}. Cooling time isochrones of $0.1$, $0.5$, $1$, $2$ and $5$ Gyrs are shown both for CO and ONe white dwarf models with dashed and solid black lines, respectively.  Light-blue dashed and solid lines indicate the crystallization onset for CO and ONe white dwarf models, respectively.
Crosses indicate the
location of observed ultra-massive white dwarfs in the literature. 
This Figure reveals that ONe-core white dwarfs are more compact than their CO-core counterparts with the same mass, specially for the most massive models. Further, the isochrones show that ONe-core white dwarfs evolve significantly faster. Therefore, the core-chemical composition of an ultra-massive white dwarf is crucial when determining its mass and its cooling times using  $\log \rm g$ and $\rm T_{eff}$ measurements. Assuming erroneously that an ultra-massive white dwarf has CO-chemical composition, would lead to an overestimation of its mass and and incorrect determination of its cooling age. 
The change of slope in the isochrones reflects the fact that, at early stages of their evolution, higher mass white dwarfs evolve faster, while the opposite occurs at advances stages.
 Finally, crystallization occurs at lower effective temperatures in CO white dwarfs.

\subsection{The colors of our ultra-massive CO models}
\label{ss:GAIA} 

We provide absolute magnitudes for our evolving models in {\it Gaia} and SDSS passbands, on the basis of the atmosphere models of \cite{2010MmSAI..81..921K,2019A&A...628A.102K}. We considered pure H atmosphere models for our H-rich sequences, and pure He atmosphere models for our H-deficient sequences.

\begin{figure}
      {\includegraphics[width=1.0\columnwidth, clip=true,trim=0 0 0 0]{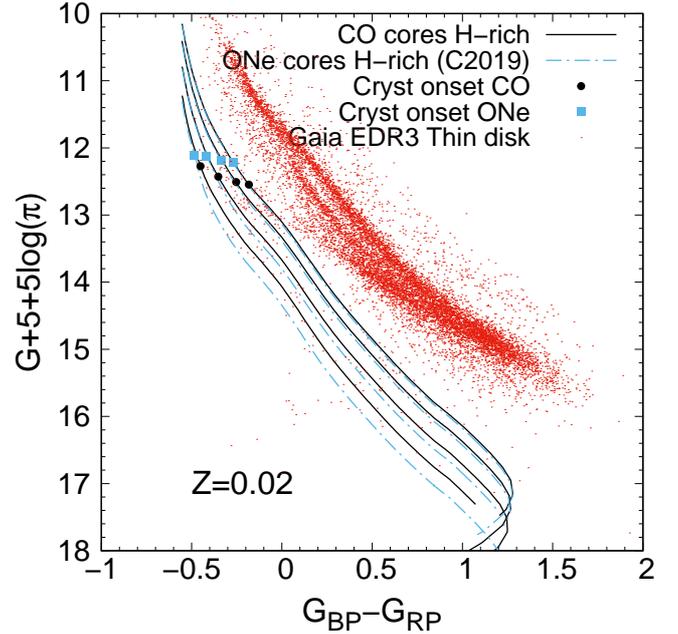}}
  \caption{{\it Gaia} color-magnitude diagram of the thin disk population of the EDR3 white dwarf sample within 100\,pc (red dots). We show our theoretical cooling tracks for ultra-massive H-rich CO white dwarfs as solid black lines, together with the ONe-core, ultra-massive H-rich theoretical cooling tracks of \protect \cite{camisassa2019} as doted-dashed cyan lines. The crystallization onset on CO models is marked using black circles, whereas the crystallization onset for ONe models is marked using cyan squares. Each set of cooling tracks contains four different masses. The masses are, from top to bottom, 1.10, 1.16, 1.23 and 1.29 $\rm M_\odot$, respectively.} 
  \label{f:GAIA}
\end{figure}

The ultra-massive H-rich CO-core cooling sequences calculated in this work are shown using solid black lines in the {\it Gaia} color magnitude
diagram in Figure \ref{f:GAIA}. For illustrative purposes we also show
the H-rich
ONe sequences with the same masses of \cite{camisassa2019} using dot-dashed cyan lines,
together with the 100 pc thin disk white dwarf population of the {\it Gaia} Early Data Release
3 (Torres et al., in preparation). The black circles (cyan squares) indicate the crystallization onset
in each CO (ONe) evolutionary sequence.
All sequences evolve from brighter magnitudes to fainter magnitudes as the evolution proceeds, but ONe white dwarfs evolve faster.
In this Figure we can see that ultra-massive white dwarfs are fainter objects than the vast majority of white dwarfs in the Galaxy, due to their smaller characteristic radius. Moreover, ultra-massive ONe white dwarfs are more compact than their CO counterparts and, therefore, they are even fainter. This should be carefully taken into account when using the photometric properties of ultra-massive white dwarfs to determine their masses and cooling ages, since considering CO and ONe models yields different results. 
Finally, the crystallization process occurs at higher luminosities in ONe white dwarfs than in CO white dwarfs. Therefore, the impact of crystallization
on the white dwarf cooling times is less important in 
an ONe white dwarf than in a CO white dwarf with the same mass.
As previously noticed in \cite{2020ApJ...902...93B}, our models confirm that the so-called Q branch is located where CO crystallization is expected to occur, not where ONe crystallization takes place. 

\section{Conclusions}

We present the first set of ultra-massive CO white dwarf cooling sequences in the literature that considers
a realistic initial chemical profile, the energy released by latent heat and phase separation during crystallization, and the energy released by the slow sedimentation of $^{22}$Ne. The initial chemical profiles are taken from the reduced mass loss scenario in \cite{ALTUMCO2021}.
A total of 32 cooling sequences have been calculated in this work.
We considered four different initial metallicities $Z=0.001$, $Z=0.02$, $Z=0.04$ and $Z=0.06$. This is, for each initial metallicity, we set the initial $^{22}$Ne abundance to be constant and equal to the value of $Z$.
For each metallicity, four white dwarf masses were considered: $1.10\, \rm M_\odot$, 
 $1.16\, \rm M_\odot$, $1.23\, \rm M_\odot$ and $1.29\, \rm M_\odot$, and we calculated both H-rich and H-deficient sequences for each mass.
 We have not explored the evolution of white dwarfs more massive than  $1.29\, \rm M_\odot$ because, for such massive white dwarfs, general relativistic effects on the structure and on the energetic become important. We defer the exploration of relativistic evolutionary models to a future work (Althaus et al., in preparation). We provide magnitudes in {\it Gaia} and SDSS passbands for all these models and made them publicly available. 
 
 Ultra-massive white dwarfs harbouring ONe cores evolve significantly faster than their CO counterparts. The fastest cooling in ONe white dwarfs is mainly due to three reasons. First, the thermal content in ONe white dwarfs is smaller. Second, crystallization occurs at higher luminosities in ONe white dwarfs, being its impact on the cooling times less important. And third, $^{22}$Ne sedimentation is a less efficient process in an ONe plasma than in a CO plasma. Further, ONe-core white dwarfs are more compact than CO-core white dwarfs with the same mass. 
 These properties should be carefully taken into account when using the photometric and spectroscopic properties of ultra-massive white dwarfs to determine their masses and cooling ages, since considering CO and ONe models yields different results. 
 
 We found that the evolution of CO white dwarfs in high-metallicity environments should be strongly affected by
 $^{22}$Ne sedimentation. These white dwarfs evolve several Gyrs slower than CO white dwarfs coming from solar metallicity progenitors. Therefore, the initial $^{22}$Ne
content in CO ultra-massive white dwarfs is crucial to determine accurate white dwarf cooling times. 

 We compared our ultra-massive cooling models with the models of \cite{Bedard2020} with the same mass, and found that our models cool down slower.  This is because
 our models consider the energy released by phase separation due to crystallization and by $^{22}$Ne sedimentation, while the models of \cite{Bedard2020} neglect these energy sources, and possibly small differences in the cooling times arise due to the difference in the chemical profiles. The cooling curves of \cite{Bedard2020} resemble our cooling sequences in which the energy released by $^{22}$Ne sedimentation is negligible ($Z=0.001$).

This set of cooling sequences, together with the cooling ONe tracks of \cite{camisassa2019} provides a powerful tool to study the ultra-massive white dwarfs in our Galaxy. 
 A considerable fraction of the ultra-massive CO white dwarf population can also be formed as a result of merger episodes, and, in this case, their $^{22}$Ne content could be higher than $0.02$ \citep{2012ApJ...757...76S}. This set of cooling sequences can be used to study the ultra-massive white dwarf population born in stellar mergers, although the cooling times obtained for double white dwarf merger models where a complete mixing of the components occurs could be slightly larger than the cooling times presented in this paper, due to differences in the chemical profiles
\citep{2014MNRAS.438...14D,ALTUMCO2021}.
Finally, we wish to remark that the core-chemical composition of ultra-massive white dwarfs is not clear to date. There are evolutionary calculations of the progenitor stars of ultra-massive white dwarfs in the literature that predict that a CO composition may be achieved \citep{ALTUMCO2021,2007MNRAS.380..933Y}, and others that predict ONe composition \citep{2006A&A...448..717S,2021ApJ...906...53S}. Further, it has not been possible to distinguish the core-chemical composition of ultra-massive white dwarfs from their observed properties, although a promising avenue is through Asteroseismology \citep{2019A&ARv..27....7C,2019A&A...632A.119C,
2021A&A...646A..60C}.
Therefore, these cooling sequences must be considered together with the ONe models of \cite{camisassa2019} as a full set of ultra-massive white dwarf cooling models. We hope that these models will help to shed light on the core-chemical composition of ultra-massive white dwarfs.

\label{conclusions}

\section{Supplementary material}

The white dwarf evolutionary models presented in this paper are available for downloading. There are two sets of tables, one with hydrogen-rich models and one with hydrogen-deficient models. Each set contains 16 ultra-massive CO-core white dwarf models, with four
different masses: 1.10, 1.16, 1.23 and 1.29 $\rm M_{\sun}$
and four different $^{22} \rm Ne$ abundances: 0.001,
0.02, 0.04 and 0.06. The $^{22} \rm Ne$ abundance roughly
corresponds to the initial metallicity of the star.
Further, we have included the 8 ultra-massive ONe-core
white dwarf models of \cite{camisassa2019}. The columns
listed are, from left to right,  effective temperature
(in K), logarithm of the luminosity (in solar units),
cooling time (in $10^{9}$years), logarithm of the surface
gravity (in CGS), stellar radius (in solar units),
magnitudes in the Sloan Digital Sky Survey passbands
($u$, $g$, $r$, $i$ and $z$) and magnitudes in the
{\it Gaia} DR2 and EDR3 passbands ( G2, $\rm G_{BP}2$,
$\rm G_{RP}2$,  G3, $\rm G_{BP}3$, and $\rm G_{RP}3$).

\section*{Acknowledgements}
We thank the comments of our expert referee, P.-E. Tremblay that helped us
to improve the original version of this work.
This work  was partially supported by the NASA 
grants 80NSSC17K0008 and 80NSSC20K0193 and the University of Colorado Boulder.
This work has made use of data from the European Space Agency (ESA) mission {\it Gaia} (\url{https://www.cosmos.esa.int/gaia}), processed by the {\it Gaia} Data Processing and Analysis Consortium (DPAC, \url{https://www.cosmos.esa.int/web/gaia/dpac/consortium}). Part of this work was supported by AGENCIA through the Programa de Modernizaci\'on Tecnol\'ogica BID 1728/OC-AR, by the PIP 112200801-00940 grant from CONICET,  by grant G149 from University of La Plata, and by the Spanish project {\it PID 2019-109363GB-100}. ST acknowledges the support by the MINECO grant AYA-2017-86274-P and by the AGAUR (SGR-661/2017).

\section*{Data Availability Statement}

Supplementary material will be available to all readers.
The cooling sequences are publicly available for download at
\href{ http://evolgroup.fcaglp.unlp.edu.ar/TRACKS/tracks.html}{http://evolgroup.fcaglp.unlp.edu.ar/TRACKS/tracks.html}.



\bibliographystyle{mnras}
\bibliography{UMCO.bib}


\bsp	
\label{lastpage}
\end{document}